# Nonvolatile ferroelectric control of ferromagnetism in (Ga,Mn)As

I. Stolichnov[1], S.W.E. Riester[1], H.J. Trodahl[1,2], N. Setter[1],

A.W. Rushforth[3], K.W. Edmonds[3], R.P. Campion[3], C.T. Foxon[3], B.L. Gallagher[3] and

T. Jungwirth[4,3]

[1] *Ceramics Laboratory, EPFL-Swiss Federal Institute of Technology,*

*Lausanne 1015, Switzerland*

[2] *MacDiarmid Institute for Advanced Materials and Nanotechnology, Victoria*

*University, Wellington, New Zealand*

[3] *School of Physics and Astronomy, University of Nottingham, Nottingham NG7 2RD,*

*United Kingdom*

[4]*Institute of Physics ASCR v.v.i., Cukrovarnická 10, 162 53 Praha 6, Czech Republic*

There is currently much interest in materials and structures that provide coupled ferroelectric and ferromagnetic responses, with a long-term goal of developing new memories and spintronic logic elements[1-4]. Within the field there is a focus on composites coupled by magnetostrictive and piezoelectric strain transmitted across ferromagnetic-ferroelectric interfaces[5-7], but substrate clamping limits the response in the supported multilayer configuration favoured for devices[1, 8]. This constraint is avoided in a ferroelectric-ferromagnetic bilayer in which the magnetic response is modulated by the *electric field* of the poled ferroelectric. Here, we report the realization of such a device using a diluted magnetic semiconductor (DMS) channel and a polymer ferroelectric gate. Polarization reversal of the gate by a single voltage **pulse results in a persistent modulation of the Curie temperature as large as 5%. The device demonstrates direct and quantitatively understood electric-field-mediated coupling in a multiferroic bilayer and may provide new routes to nanostructured DMS materials and devices via ferroelectric domain nano-patterning. The successful implementation of a polymer-ferroelectric gate field-effect transistor (FeFET) with a DMS channel adds a new functionality to semiconductor spintronics and may be of importance for future low-voltage spintronics devices and memory structures.**

Non-volatile electric-field control of ferromagnetism using a ferroelectric gate has been reported in oxide ferromagnetic layers that do not lend themselves to integration with semiconductors. In particular, such control has been found in manganites[9] and Co doped $TiO_2$[10], although in the latter case it seems probable that the material is phase segregated and the ferromagnetism arises from Co clusters[11]. In contrast, here we utilize the well-established (Ga,Mn)As DMS as the conducting channel in the FeFET structure, which offers close compatibility with its parent compound semiconductor GaAs. In this system Mn substituting for Ga is an acceptor providing both local magnetic moments and valence band holes[12]. The control of ferromagnetism using the ferroelectric gate demonstrated here relies on the mediation of the Mn-Mn exchange interaction by the strongly spin-orbit coupled valence band holes which control both the strength of the magnetic interactions and the magnetocrystalline anisotropies[12]. The Curie temperature $T_C$ we focus on in this paper can thus be a significant function of the hole density p, offering the potential for altering the ferromagnetic response by electric-field control.

Control of the magnetic response in a DMS channel of a conventional FET was first reported by Ohno et al. [13] and subsequently similar results in magnetically-doped group IV, III-V and II-VI semiconductors have been reported[14-17]. In such devices, control requires the application of a large gate voltage and is not persistent. In contrast a ferroelectric gate offers the potential for the large non-volatile carrier-density control needed in these heavily doped materials, by modest voltages (less than 5 V in ultra-thin ferroelectric films). Ferroelectric gates can offer sub-nanosecond response times, and nanometer-scale ferroelectric domain engineering to modulate magnetic properties with rewritable ferroelectric domain patterns.

Achieving a FeFET based on magnetic (Ga,Mn)As presents a number of challenges. The change in 2D hole concentration that can be produced by a ferroelectric gate does not exceed the spontaneous polarization of the ferroelectric material, typically about 10 μC cm$^{-2}$ (6x10$^{13}$ cm$^{-2}$) and screening at the ferroelectric-semiconductor interface can reduce the induced charge below that ideal. The 3D hole density in ferromagnetic (Ga,Mn)As is high (10$^{20}$-10$^{21}$ cm$^{-3}$ (10$^{13}$-10$^{14}$ cm$^{-2}$ per nm thickness)). One therefore requires the thinnest possible (Ga,Mn)As layers. Thus we have implemented our device with the thinnest (7 nm) layers of 6 % Mn-doped GaAs that exhibit consistent and well-characterized ferromagnetic behavior. A second challenge relates to an incompatibility between (Ga,Mn)As and the 400-600 °C anneal required for oxide ferroelectrics. A temperature much above 250 ºC leads to a loss of substitutional Mn, reducing both the local moment density and the hole concentration on which the Mn-Mn exchange relies[18].

Thus, instead of a conventional perovskite ferroelectric, we have used a co-polymer polyvinylidene fluoride with trifluoroethylene P(VDF-TrFE), which requires only a 140 ºC anneal to align the polymer chains[19].

A cross section of the FeFET structure is shown in Fig. 1. The 7 nm (Ga,Mn)As films were grown by low temperature ($\leq 250$ ºC) molecular beam epitaxy onto a 330 nm high-temperature GaAs buffer layer on a semi-insulating GaAs (001) substrate. Two films were tested; film (I) shown in Fig.1 was separated from the buffer by a 50 nm thick $Al_{0.33}Ga_{0.67}As$ barrier, whereas film (II) was deposited directly onto the buffer layer. Films I and II have room temperature sheet resistances of 11 and 8 k$\Omega$ respectively.

Hall bars required for magnetotransport studies were defined by chemical etching and Ti/Au (15 nm/ 125 nm) contacts were deposited by e-beam evaporation. The 200 nm P(VDF-TrFE) ferroelectric gate polymer was spin-coated from the 2% methyl ethyl ketone solution and crystallized for 10 minutes at 137°C. Finally a 100 nm Au gate electrode was deposited by thermal evaporation.

The ferroelectric P(VDF-TrFE) films on (Ga,Mn)As show the sharp polarization hysteresis loop seen as an inset in Fig. 2a, with an ambient-temperature remnant polarization close to 6 $\mu C/cm^2$. The polarization is stable for more than a week and after repeated cycling to 10 K, as confirmed by both polarization hysteresis and transverse piezoelectric response measured directly through the gate electrode by piezoelectric force microscopy[20] at room temperature. In our magneto-transport experiments the gate was poled with a single +30V (-30V) voltage pulse.

Ferroelectric control of the hole density in the channel can be seen in Fig. 2a, which compares the temperature-dependent resistivities of the channel in film (I) before and after poling the gate and a reference Hall bar with the ferroelectric gate removed. Polarizing the gate induces a resistivity change between accumulation and depletion of ~4% at 300 K, ~9% around $T_C$ and ~11% at 20 K. For all three curves, the resistance initially increases as the temperature is lowered, before reaching a maximum below which it decreases. This resistance maximum is generally found to occur close to the Curie temperature, $T_C$, and the fall in resistance below $T_C$ is attributed to suppression of spin-disorder scattering[18]. The position of the peak (Fig. 2a, lower inset) changes from 91 K in accumulation to 85 K in depletion with that of the reference bar between these values. This demonstrates that the polarization state of the gate significantly modulates the Curie temperature.

A higher degree of accuracy in inferring the Curie temperatures of (Ga,Mn)As materials can be achieved from the measured Hall resistivity, $\rho_{xy}$, which is dominated by the extraordinary Hall resistivity[21]. The ratio ($\rho_{xy}/\rho_{xx}^n$), where $\rho_{xx}$ is the longitudinal resistivity and n~2, is expected to track the magnetization (M) for metallic materials, in the Berry phase interpretation of the extraordinary Hall resistivity [22]. One can then obtain $T_C$ values from Arrott plots [$((\rho_{xy}/\rho_{xx}^2))^2$ vs $B/(\rho_{xy}/\rho_{xx}^2)$], where B is magnetic induction, since ferromagnetic (paramagnetic) states correspond to positive (negative) intercepts on the ordinate of the extrapolated linear behaviour at higher B values [21, 23]. The Arrott plots, for n=2, in accumulation and depletion, show clear shifts in the ordinate intercepts (Fig. 3a,b). This intercept is plotted vs. T in Fig. 3c, from which $T_C$ can then be read as the temperature at which the intercept passes through zero. This gives a Curie temperature which is ~3.8 K (4.7 %) lower in depletion than in accumulation (Fig. 3c). Since the magnetic field dependence of $\rho_{xy}$ is much larger than that of $\rho_{xx}$ for our samples the shift in $T_C$ obtained in this way is barely sensitive to the value of n: taking n=1 ( corresponding to the skew scattering picture[22]) gives a $T_C$ shift of ~3.2K (see Fig. 3d).

We now turn to the second structure, with a channel consisting of a (Ga,Mn)As film (II) without an (AlGa)As hole barrier. Figure 4a displays the temperature-dependent resistivities in the accumulation and depletion states and for a reference Hall bar after removing the ferroelectric gate. The shift of the resistance peak and the Arrott plot intercepts (Figure 4b) both indicate a control of $T_C$, with an intermediate resistance maximum and $T_C$ in the Hall bar with the gate removed. The observed $T_C$ shift of 1.8 K (2.2 %) is weaker than in film (I). This may in part be due to a higher carrier concentration in film II consistent with the lower resistivity and/or the stronger confinement of holes (and hence smaller effective thickness) in the (Ga,Mn)As film I grown on the wider bandgap (Ga,Al)As material.

Table I summarizes the data on $T_C$ shift driven by the ferroelectric gate. These measured $T_C$ shifts are in good agreement with the predicted sub-linear hole-concentration dependence $T_C(p)$[24]. From Table I the $T_C$-shift associated with the resistance modulation in film I (II) gives $\partial(\ln T_C)/\partial(\ln R) \approx -0.5\ (-0.7)$. A detailed comparison can be made with the dependence on hole concentration of $T_C$ and the conductivity as calculated microscopically within the kinetic-exchange model band structure and Boltzmann transport theory [25-27]. Within the model a 6 % Mn-doped film reaches $T_C$ of 80 K for a hole concentration of $0.25 \times 10^{21}$ cm$^{-3}$. Computed Curie temperatures for varying hole concentrations then indicate that $\partial T_C/\partial p = 2.15 \times 10^{-19}$ K cm$^3$ and $\partial(\ln T_C)/\partial(\ln p) = 0.68$. The calculated conductivity ($\sigma \alpha\ 1/R$) is super-linear with $\partial(\ln R)/\partial(\ln p) = -1.4$, which then predicts $\partial(\ln T_C)/\partial(\ln R) = -0.48$, in agreement with our experimental data. Alternative Kubo formula calculations based on *ab initio*, coherent potential approximation band structures[28] predict $\partial(\ln T_C)/\partial(\ln R) \sim -0.6$ for the relevant Curie temperatures and resitivities, again in reasonable agreement with our experiments.

Hysteretic behavior of the magnetisation is another important signature of ferromagnetism and its change associated with the switching between accumulation and depletion states clearly indicates that the magnetic state of the system has been altered. Figure 5 shows hysteresis loops of Hall resistance of film (I) plotted as a function of magnetic field oriented along the Hall bar axis, which corresponds approximately to the easy magnetization axis. Hysteresis is observed when the magnetisation rotates through 180°. We are able to observe this switch of magnetisation as a change in Hall resistance, due to either the planar Hall effect[29], or a very slight off-plane orientation of the magnetic easy axis (about 0.1°) giving an anomalous Hall signal. The difference in the switching field is clearly observed throughout the temperature range and becomes more pronounced when approaching $T_C$. The hysteresis cannot be resolved at 60K in the depletion state and above 64K in the accumulation state consistent with a $T_C$ difference of approximately 4K.

The results reported here establish a number of important advances toward implementation of an FeFET multiferroic device, including quantitative description based on theoretically-predicted and experimentally-determined behaviour of the magnetic properties of (Ga,Mn)As. A widely-recognised incompatibility between ferroelectrics and III-V semiconductors has been solved by integration of a polymer ferroelectric in an FeFET configuration. Poling and retention in the polymer ferroelectric gate and control of the semiconductor channel carrier density to low temperatures have all been demonstrated successfully. A polymer-gate integration in a dilute magnetic semiconductor (Ga,Mn)As FeFET has led to an electric-field-mediated control of ferromagnetism in the semiconductor channel that has been thoroughly investigated by magneto-transport measurements. The effects are in agreement with the established understanding of hole-mediated exchange in (Ga,Mn)As.

**Acknowledgements**

We acknowledge support from the Swiss National Science Foundation, Swiss Program on NanoSciences (NCCR); EU Grant IST-015728, UK Grant GR/S81407/01; CR Grants 202/05/0575, 202/04/1519, FON/06/E002, AV0Z1010052, and LC510

**Table I.** Change of sheet resistance and $T_C$ shift induced by switching ferroelectric gate

| Sample | R, kΩ, accum. | ΔR, kΩ | $T_C$ (K), accum. | $\Delta T_C$ (K) |
|---|---|---|---|---|
| I | 13.45 | 1.33 (9%) | 80.7 | 3.8 (4.7%) |
| II | 11.88 | 0.35 (2.9%) | 84.1 | 1.9 (2.2%) |

# Figure captions

Figure 1. **Schematic longitudinal cross-section of the Hall bar based on the film (I).** 7nm Mn-doped layer is separated from GaAs substrate by (Ga,Al)As barrier. The top-to-bottom direction of the spontaneous polarization P in the ferroelectric gate corresponds to the depletion state.

Figure 2. **Ferroelectric gate operation and temperature dependence of the sheet resistance for film (I)**. Hysteretic polarization reversal in the gate allows for non-volatile switching between accumulation and depletion states. Around 80K the sheet resistance in the depleted state is 9% higher than in the accumulation state and the reference sample with ferroelectric gate removed lies in between.
**Upper inset:** Polarization hysteresis loop of the gate driven with 1KHz triangle pulse.
**Lower inset:** Curves of resistance vs. temperature normalized and zoomed in the region around the maximum occurring close to $T_C$. In the depletion state the maximum shifts towards lower temperatures as expected.

Figure 3. **Control of ferromagnetism via ferroelectric switching: determination of the ferromagnetic $T_C$ change by Arrott plots for film (I).** Arrott plots for the accumulation (a) and depletion (b). (c) The intercepts from Arrott plots plotted vs. temperature, which deliver $T_C$ of 76.9K and 80.7K, respectively, for depletion and accumulation. (d) Alternative Arrott plots with $[((\rho_{xy}/\rho_{xx}))^2 \text{ vs } B/(\rho_{xy}/\rho_{xx})]$ corresponding to skew scattering delivers similar $T_C$ shift.

Figure 4. **Temperature dependence of the sheet resistance and control of ferromagnetism for the structure based on film (II)**.

(a) Curves of sheet resistance of the film (II) vs. temperature show weaker difference between accumulation and depletion states compared to the film (I).

**Upper inset:** Polarization hysteresis loop of the gate driven with 1KHz triangle pulse.

**Lower inset:** Curves of resistance vs. temperature normalized and zoomed in the region around the maximum.

(b) Determination of the Tc from Arrott plot data from film (II) with intercept plotted vs. temperature as in Fig 3c.

Figure 5. **Control of ferromagnetism via ferroelectric switching: change of hysteretic properties by poling ferroeletric gate.** The difference in the coercive field is clearly observed in the hysteresis loops of Hall resistance of film (I) plotted vs. magnetic field for accumulation and depletion states for 50K (a), 56K (b) and 60K (c). At 60 K the hysteresis cannot be resolved in the depletion state, while in the accumulation state it is still clearly observed.

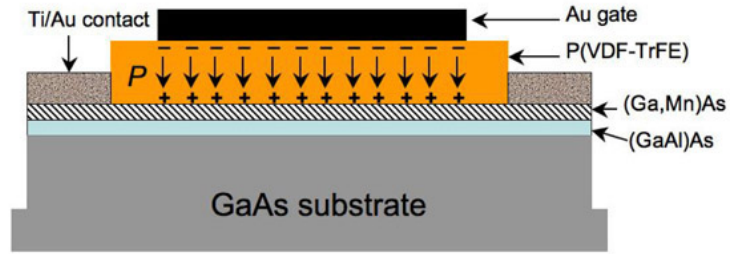

Figure 1

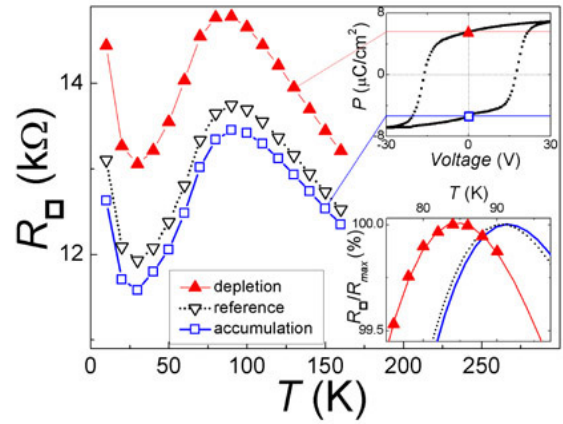

**Figure 2**

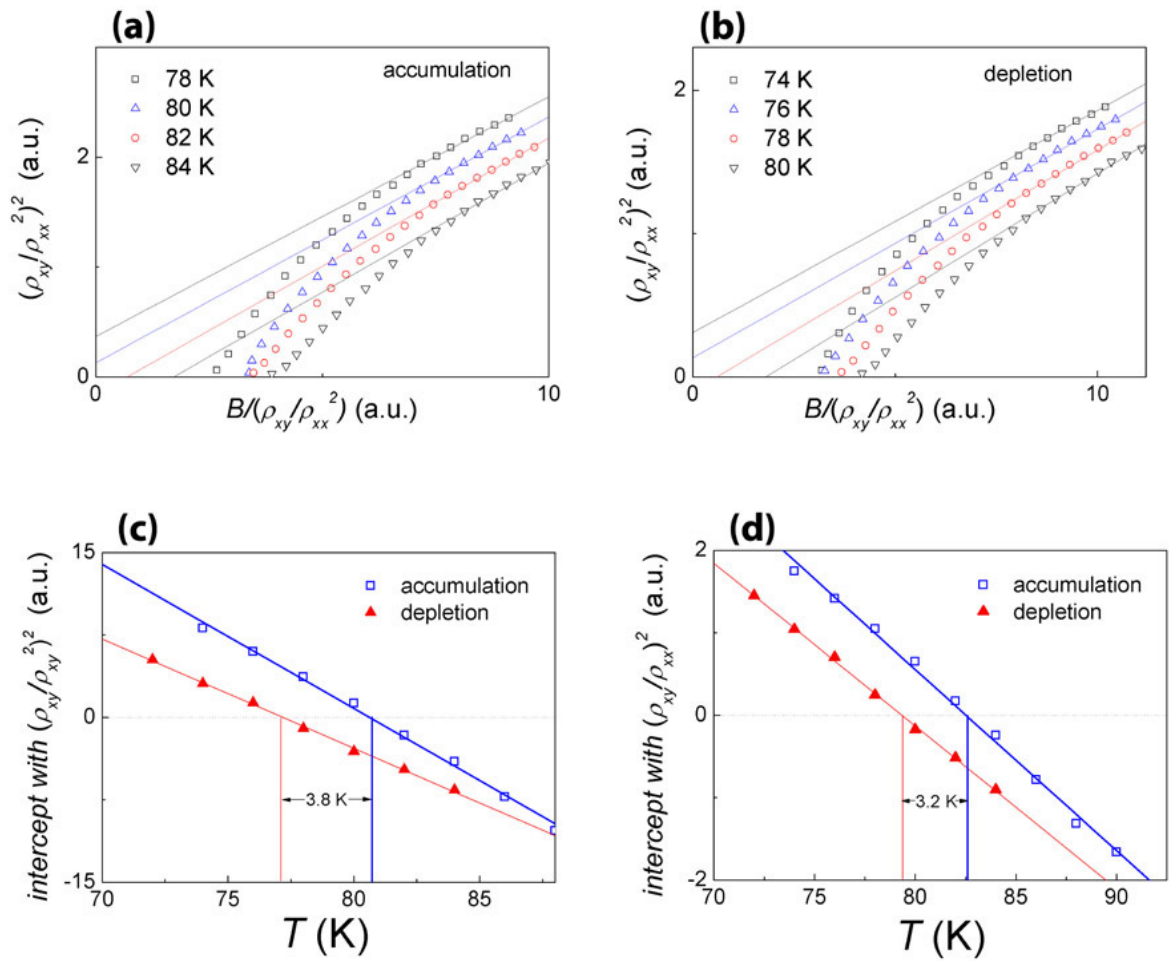

**Figure 3**

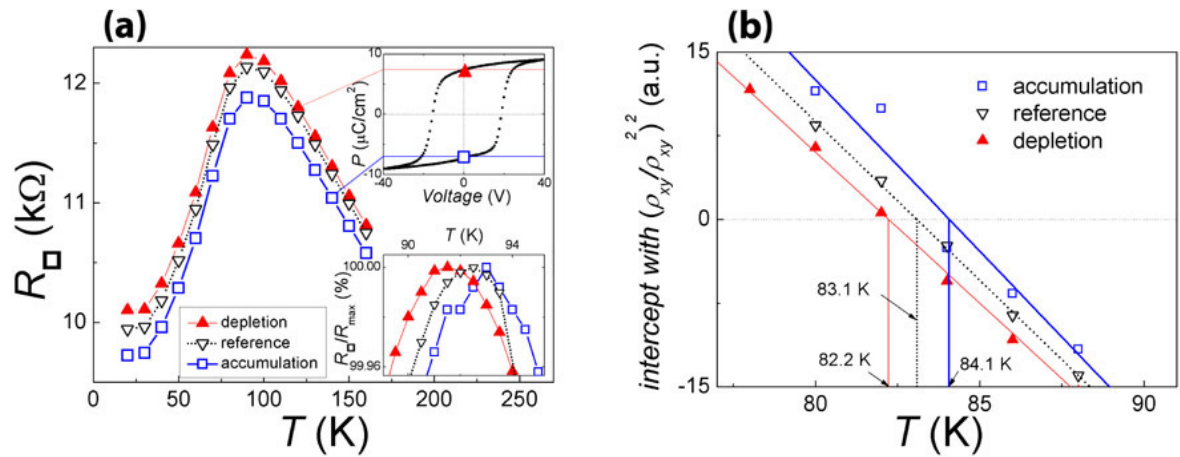

**Figure 4**

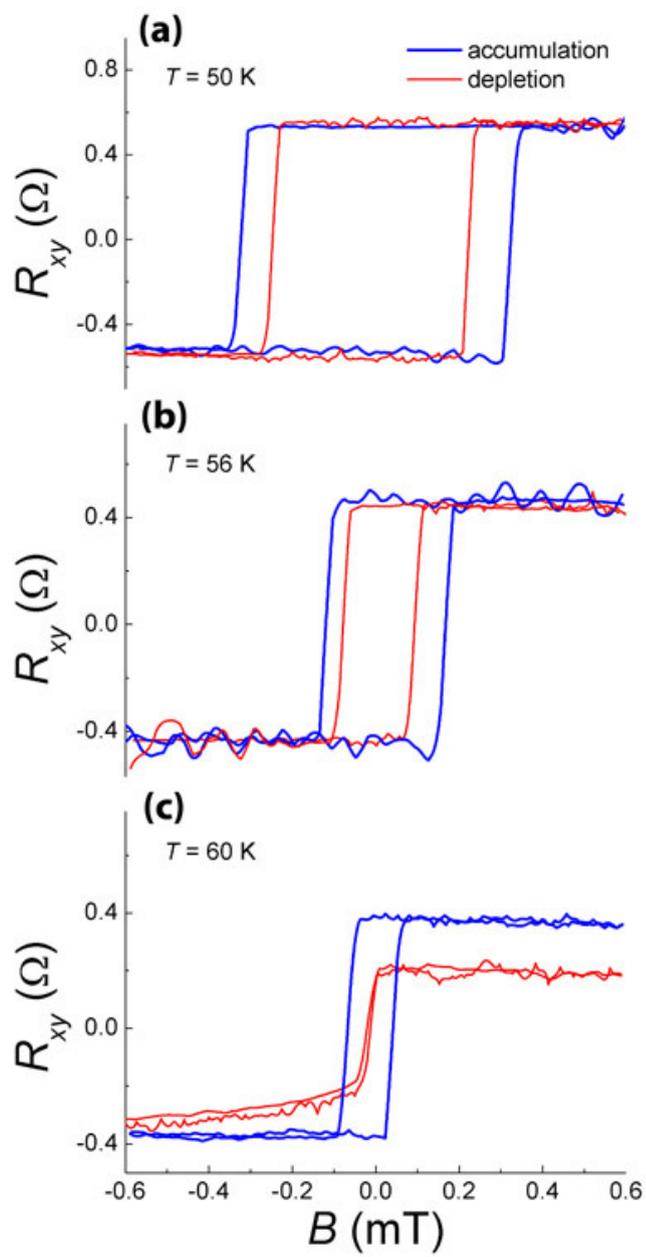

**Figure 5**